\documentclass[12pt,a4paper,final]{iopart}

\renewcommand{\figurename}{Figure}
\expandafter\let\csname equation*\endcsname\relax

\expandafter\let\csname endequation*\endcsname\relax
\usepackage{graphicx}
\usepackage{dcolumn}
\usepackage{bm,soul}
\usepackage{color}
\usepackage{amssymb,amsmath,physics}

\begin{document}
 
\title{Exact statistics and thermodynamic uncertainty relations for a periodically driven electron pump}

\author{Pedro E. Harunari$^{1}$, Carlos E. Fiore$^{1}$ and Karel Proesmans$^{2,3}$}
\address{$^{1}$ Institute of Physics of S\~ao Paulo University, 
Rua do Mat\~ao, 1371, 05508-090
S\~ao Paulo, SP, Brazil\\$^{2}$ Simon Fraser University, 8888 University Drive, Burnaby, British Columbia, Canada\\$^{3}$ Hasselt University, B-3590 Diepenbeek, Belgium\\
}
\ead{pedro.harunari@usp.br}
\date{\today}

\begin{abstract}
We introduce a model for a periodically driven electron pump
  that sequentially interacts with an arbitrary number of heat and particle reservoirs. Exact
  expressions for the thermodynamic fluxes, such as
  entropy production and particle flows are derived arbitrarily far
  from equilibrium. We use the present model to perform a comparative study of thermodynamic uncertainty relations that are valid for systems with time-periodic driving.
\end{abstract}

\maketitle

\section{\label{intro}Introduction}
Over the last few decades, stochastic thermodynamics has emerged
as a theory to describe the thermodynamic properties of mesoscopic systems
using Markov dynamics \cite{seifert2012stochastic,van2015ensemble}.
It   not only reproduces the fundamental
results from classical thermodynamics,  but also introduces several
new concepts, such as the Jarzynski equality \cite{jarzynski1997nonequilibrium,Jarzynski2011}, the thermodynamics of information processing \cite{parrondo2015thermodynamics}, the thermodynamics of phase
transitions \cite{noa2019entropy, herpich2018collective, herpich2019universality, campisi2016power, goes2020quantum}, an extension of linear irreversible thermodynamics
to time-periodic systems \cite{brandner2015thermodynamics, proesmans2015onsager, proesmans2016linear, brandner2016periodic, PhysRevE.100.022141} and others.
Furthermore, technological advancements have opened up the
possibility to probe these theoretical predictions on real microscopic systems \cite{martinez2016brownian,proesmans2016brownian,ciliberto2017experiments,josefsson2018quantum}.

Recently, a general bound, known as the thermodynamic uncertainty relations (TUR) has been derived \cite{barato2015universal,pietzonka2016universal,gingrich2016dissipation,seifert2018stochastic,proesmans2018case,horowitz2019thermodynamic} {following its proposal in the context of biomolecular processes \cite{barato2015thermodynamic}}.
It shows for steady-state Markov systems that the signal to noise ratio of any thermodynamic flux is bounded by half the entropy production rate. Such a relationship has been used to infer the entropy production rate \cite{gingrich2017inferring,li2019quantifying,manikandan2019inferring,busiello2019hyperaccurate} and to bound the performance of mesoscopic heat engines \cite{pietzonka2018universal}. Furthermore, it has been verified experimentally \cite{hwang2018energetic,pal2019experimental}, linked to the fundamental symmetries of the system \cite{guioth2016thermodynamic,hasegawa2019fluctuation,vroylandt2019isometric}, and  several extensions to other thermodynamic observables have been derived \cite{dechant2018current,di2018kinetic}. One can however show that the original TUR does not hold for systems with time-periodic driving \cite{barato2016cost,ray2017dispersion,holubec2018cycling}. Several extensions of the original TUR have been proposed  \cite{proesmans2017discrete,barato2018bounds,koyuk2018generalization,proesmans2019hysteretic,barato2019unifying,koyuk2019operationally}, raising the question of how tight they are and how they perform in different regimes. Such a quantitative comparison has however not been done.

One of the drawbacks concerning the study of periodically driven systems is that the number of exactly solvable models is rather limited \cite{engel2009asymptotics,nickelsen2011asymptotics,ryabov2013work,holubec2014exactly}. One notable exception is the electron pump model, derived by Rosas et al. \cite{rosas1,rosas2}. In this model, a two-level system is 
 sequentially brought into contact with two or three reservoirs, which allows for the transfer of electrons between those reservoirs.  Due to the simplicity of a two-level system, it is possible to derive analytic results \cite{vsubrt2007exact,chvosta2010thermodynamics,verley2013modulated}. Here, we extend those results to a model with an arbitrary number of reservoirs.

One of the strengths of this model is that one has
full access to all relevant quantities, such as number of thermodynamic fluxes, their affinities and the
driving period. Therefore,
it is possible to look at several regimes, such as the near
equilibrium regime (where the thermodynamic fluxes can be determined
from the Onsager coefficients) and the regime where one of the affinities is much larger than the others.
This flexibility makes the system the perfect toy model to do a
quantitative comparison of the time-periodic thermodynamic uncertainties relations.

This paper is organized as follows. In Sec.~\ref{model} we describe the model and show how the solutions from \cite{rosas1,rosas2} can be extended to an arbitrary number of reservoirs.
These results are used in Sec.~\ref{stother} to study the stochastic thermodynamics of our model. In Sec.~\ref{tTur}, we give an overview of the existing thermodynamic uncertainty relations for time-periodic systems. We compare those uncertainty relations in the context of the electron pump model in Sec.~\ref{compar}. Finally, our conclusions are discussed in Sec. \ref{conc}.

\section{Model and exact solution\label{model}}

\begin{figure}[h]
\includegraphics[width=.45\textwidth]{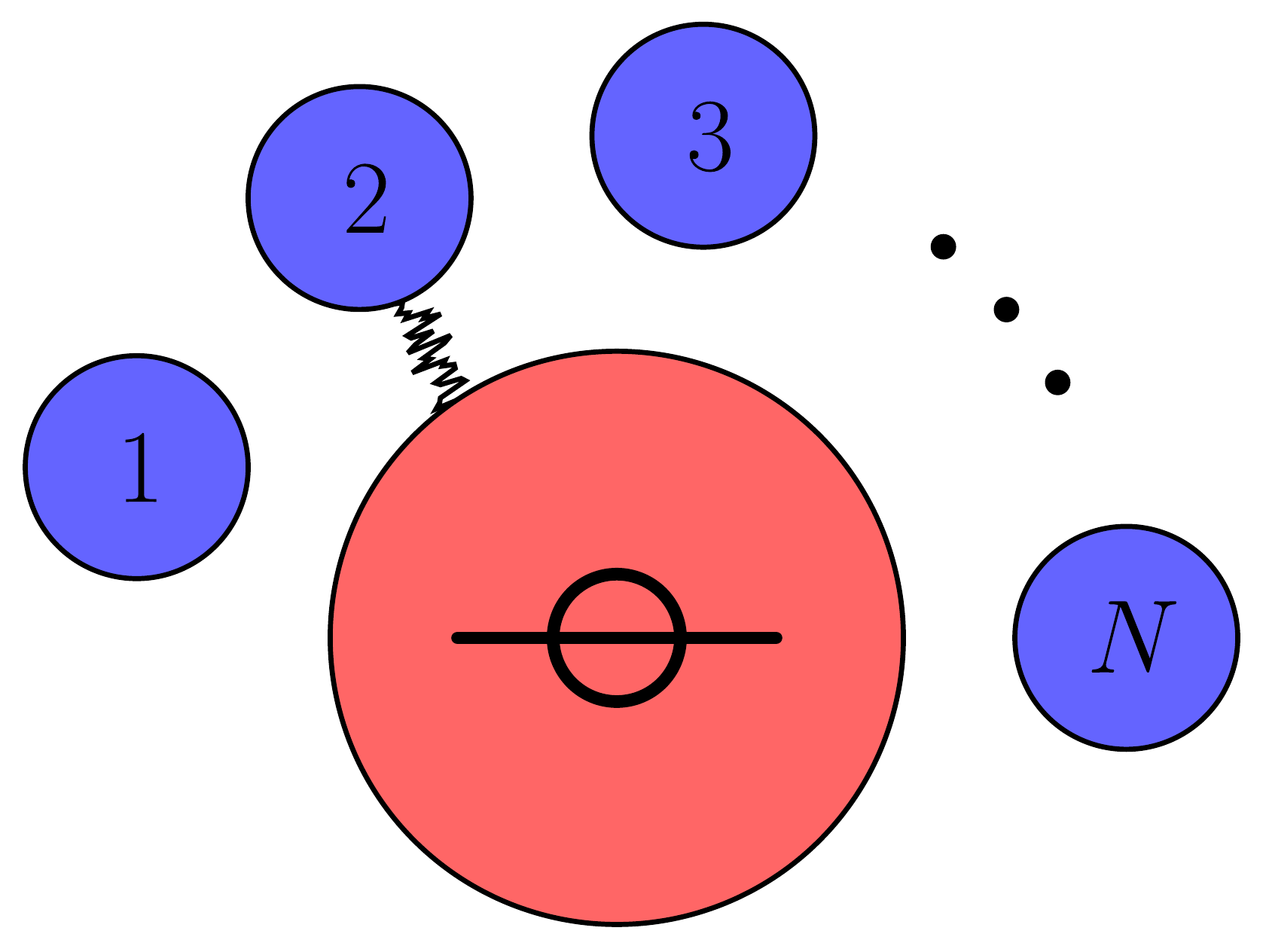}
\caption{Description of the system composed of a  quantum dot
  in contact with reservoir $2$. After an amount of time $\tau'$, the quantum dot will be decoupled from reservoir $2$ and coupled to reservoir $3$.
 }
\label{fig1}
\end{figure}

The model under study consists of
a single-level quantum dot, that can be empty or occupied by a single electron. The quantum dot is sequentially placed in contact with one of the $N$
electron reservoirs for a duration $\tau'$, after which it is decoupled from that reservoir and coupled to the next one, c.f.~Fig.~\ref{fig1}. The total
period of one cycle is $\tau=N\tau'$, after which the system
returns to its initial configuration and electrons may have
been transferred between the reservoirs.

When the quantum dot is in contact with a given reservoir,
its dynamics is described by the master equation 
\begin{equation}
  \dot{p} (t) = \omega(t) (1-p(t)) - \overline{\omega}(t) p(t),
  \label{me}
\end{equation}
where $p(t)$ denotes the probability of the quantum dot being occupied by an electron at time $t$ and { $\omega(t)$ and $\overline{\omega}(t)$ denote the transition rates for an electron to jump from the the reservoir to the system and vice versa. Throughout this paper we shall assume that the quantum dot is always in contact with a single reservoir with constant jump rates, i.e., jump rates are
$\omega(t)=\omega_i$ and $\overline{\omega}(t)=\overline{\omega}_i$ when $t\in [(i-1)\tau' , i\tau'[$ and $i\in\{1,...,N\}$. }

{ Since two-level systems coupled to a single reservoir relax to an  equilibrium state,  the probability of the quantum dot being occupied by an electron  when it is in contact with only one reservoir $i$ converges to}
\begin{equation}
  p^{eq}_i{ \equiv} \frac{\omega_i}{\omega_i+\overline {\omega_i}}.
\end{equation}
This will no longer be true when the quantum dot is periodically connected to different reservoirs for a finite amount of time. In that case, the reservoirs exchange electrons and the system continuously produces entropy.
One can however find the occupation probability by solving the master equation Eq.~(\ref{me}) explicitly, which leads to {
\begin{equation}\label{formalsol}
	 p(t) = p_i^\text{eq} + \left[p \left((i-1)\tau'\right) - p_i^\text{eq}\right] e^{-(\omega_i + \overline{\omega}_i) (t-(i-1)\tau')},
\end{equation}
{where the value of the integer $i$ in the right-hand side is the one which satisfies $(i-1)\tau' \leq t < i\tau'$, i.e., it is the index of the reservoir that is interacting with the quantum dot at time $t$. $p(t)$ is the exact time-periodic occupation probability of the quantum dot. With this information we can evaluate all macroscopic quantities of interest.} 
Since the system is ergodic and is driven time-periodically, it relaxes to a time-periodic steady-state distribution}  $p^\text{ss}(t+\tau)=p^\text{ss}(t)$ \cite{herrmann2015statistics}, which can be obtained solely in terms of the transition rates and the length of the period:
{
\begin{align}\label{pss}
	 p^\text{ss}(t) = p_i^\text{eq} + &e^{-(\omega_i + \overline{\omega} _i) (t-(i-1)\tau')} \times \nonumber\\ &\times \left\lbrace \frac{\xi_{1,i-1}}{1- \xi_{1,N}} \left[ \Delta_{N,1} + \sum_{n=2}^N \xi_{n,N} \Delta_{n-1,n} \right] +\sum_{m=2}^i \xi_{m,i-1} \Delta_{m-1,m} \right\rbrace,
\end{align}}
where we introduced $\Delta_{i,j} \equiv p_i^\text{eq} - p_j^\text{eq}$ and $\xi_{i,j} \equiv \exp\{ -\tau' \sum_{n=i}^j (\omega_n + \overline{\omega}_n) \}$. Throughout the paper, we will always assume that the system has reached its time-periodic steady state. We pause to make a few comments. First, a time-independent equilibrium distribution is recovered when
$\Delta$'s vanish. Second, the above general expression reduces to the
results derived in Refs.~\cite{rosas1} and \cite{rosas2} for $N=2$ and $3$ respectively. Third,
in the limit of strong coupling, $\omega_i,\overline{\omega}_i\rightarrow \infty$ for all $i$, { $p^\text{ss}(t)$} essentially  corresponds to an equilibrium distribution   $p_i^\text{eq}$.

 We shall now calculate the electron flux out of reservoir $i$. This corresponds to the difference between the occupancy  $n(t)$ at the beginning and at the end of the interaction divided by the period, ${ J}_i \equiv [n(i\tau')-n((i-1)\tau')]/\tau$, where the occupancy $n(t)=1$ if the quantum dot is occupied at time $t$ and $n(t)=0$ otherwise. By taking the average of $J_i$, we arrive at the following expression
\begin{align}
  \overline{J}_i =&\frac{p^{\textrm{ss}}(i\tau')-p^{\textrm{ss}}((i-1)\tau')}{\tau}\nonumber\\=& \frac{1}{\tau} \left\lbrace \frac{\xi_{1,i-1}}{1- \xi_{1,N}} \left[ \Delta_{N,1} + \sum_{n=2}^N \xi_{n,N} \Delta_{n-1,n} \right] +\sum_{m=2}^i \xi_{m,i-1} \Delta_{m-1,m} \right\rbrace (\xi_{i,i} - 1),
  \label{mc}
\end{align}
where the overline represents the { ensemble average.}

One can easily check that this expression satisfies the property $\sum_{i=1}^N \overline{J} _i=0$, as would be expected. Furthermore, one can also derive an expression for the variance of the flux per cycle { $\mathrm{var}(J_i) \equiv \tau\left(\overline{J^2_i} - \overline{J}_i^2\right)$}:
\begin{align}
  \mathrm{var}(J_i) =& \frac{ \overline{[n(i\tau')-n((i-1)\tau')]^2}-[\overline{n(i\tau')}-\overline{n((i-1)\tau')}]^2}{\tau}\nonumber\\
  =& \frac{1-\xi_{i,i}}{\tau}2p_i^\text{eq} (1- p_i^\text{eq} ) \nonumber\\
  &+ \frac{1-\xi_{i,i}}{\tau} (1- 2p_i^\text{eq} ) \left\lbrace \frac{\xi_{1,i-1}}{1- \xi_{1,N}} \left[ \Delta_{N,1} + \sum_{n=2}^N \xi_{n,N} \Delta_{n-1,n} \right] +\sum_{m=2}^i \xi_{m,i-1} \Delta_{m-1,m} \right\rbrace \nonumber \\
  &- \frac{(1-\xi_{i,i})^2}{\tau} \left\lbrace \frac{\xi_{1,i-1}}{1- \xi_{1,N}} \left[ \Delta_{N,1} + \sum_{n=2}^N \xi_{n,N} \Delta_{n-1,n} \right] +\sum_{m=2}^i \xi_{m,i-1} \Delta_{m-1,m} \right\rbrace^2,
  \label{var}
\end{align}
where we used the properties $\overline{n(i\tau')n((i-1)\tau')} = p^\text{ss}(i\tau'\vert n((i-1)\tau')=1) p^\text{ss}((i-1)\tau')$, $\overline{n(i\tau')} =\overline{n(i\tau')^2} =p^{\textrm{ss}}(i\tau')$ and Eqs.~(\ref{me}), (\ref{formalsol}) and (\ref{pss}).

\section{Stochastic Thermodynamics\label{stother}}
Having defined the dynamics of the system, we are now ready to introduce its stochastic thermodynamics  \cite{seifert2012stochastic,van2015ensemble,van2015stochastic}. In particular, we can define the
chemical work flux ${\dot W^\text{(chem)}}_i$ and the heat flux ${\dot Q}_i(t)$ associated with the $i$-th reservoir,
\begin{eqnarray}
 {\dot W_{i}^\text{(chem)}} { (t)}&=&\mu_i { \dot{ p}(t)},
\label{eq14-2}
\\
{\dot Q_i(t)}&=&(\epsilon-\mu_i){ \dot{ p}(t)},
\label{eq14-3}
\end{eqnarray}
for { $t$ satisfying $(i-1)\tau'\leq t<i\tau'$}. Here, we have defined the { chemical potential of the $i$-th reservoir} $\mu_i$ and the energy of the quantum dot $\epsilon$. As we are assuming the energy level of the quantum dot to be constant, the mechanical work delivered to the system will be zero. Using the definitions of the electron fluxes, one can also determine the total average delivered amount of chemical work and produced heat per cycle in the steady state:
\begin{equation}
\overline{\dot{W}}^\text{(chem)} = \sum_{i=1}^N \overline{J}_i\mu_i  = \sum_{i=2}^N \overline{J}_i(\mu_i - \mu_1) ,
\end{equation}
and
\begin{equation}
	\overline{\dot{Q}} = \sum_{i=1}^N \overline{J}_i(\epsilon - \mu_i )  = \sum_{i=2}^N \overline{J}_i(\mu_1 - \mu_i),
\label{meanheat}
\end{equation}
{ where in both cases  we used $\sum_i \overline{J}_i =0$. }

Due to the periodicity of the system, one has $\overline{\dot{Q}}=- \overline{\dot{W}}^\text{(chem)}$, in agreement with the first law of thermodynamics.

The entropy production rate, averaged over one period, can be determined using Schnakenberg's
formula \cite{schnakenberg1976network}:
\begin{equation}
\label{sch}
  \overline{\Pi}= \frac {{ 1}}{\tau} \sum_{i=1}^N \int\limits_{(i-1)\tau'}^{i\tau'} [ \omega_i - (\omega_i + \overline{\omega}_i) { p^\text{ss}(t)} ]\ln \frac{\omega_i(1-{ p^\text{ss}(t)})}{ \overline{\omega}_i { p^\text{ss}(t)}}\textrm{d} t,
\end{equation}
{ where the Boltzmann constant is set to unity here and hereafter.}

Taking into account   the periodicity of
        $p(t)$,  the mean entropy production  per unit of time, averaged over one cycle, $\overline{\Pi}$ can be rewritten to a sum of fluxes $\overline{J}_i$ and associated thermodynamic forces $X_i$
\begin{equation}
  \label{forces}
  \overline{\Pi}= \sum_{i=2}^N \overline{J}_i X_i,
\end{equation}
where  $\overline{J}_i$ is given by Eq. (\ref{mc}) and the forces $X_i$ read
\begin{equation}
  X_i\equiv  \ln \frac{\omega_i \overline{\omega}_1}{\overline{\omega}_i \omega_1}.
  \label{xidef}
    \end{equation}
This structure for the entropy production rate is a well-known result from classical non-equilibrium thermodynamics \cite{callen1998thermodynamics,de2013non}.
The first reservoir is regarded as a reference,
so the thermodynamic forces are quantified by how much they
drive the system away from the reference system, similar to \cite{rosas1,rosas2}.

By taking into account the local detailed balance assumption,
the entropy production formula can also be rewritten in terms of the heat fluxes:
\begin{equation}
    \frac{\omega_i}{\overline{\omega}_i}=e^{-(\epsilon-\mu_i)/T_i},
\end{equation}
which gives an expression for the transition rates in terms of the macroscopic variables $\mu_i$, $\epsilon$ { and the $i$-th reservoir's temperature $T_i$.}
Furthermore, the thermodynamic forces $X_i$ can now also be related to
\begin{equation}
  X_i=\frac{\epsilon - \mu_1}{T_1}-\frac{\epsilon - \mu_i}{T_i}.
  \label{xi}
\end{equation}
By combing Eqs. (\ref{meanheat}), (\ref{forces}) and (\ref{xi}), we re-obtain the classical thermodynamic definition for the mean entropy production
\begin{equation}
  \overline{\Pi}= \sum_{i=1}^N\frac{\overline{\dot{Q_i}}}{T_i}.
\end{equation}

Finally, we mention that it is also possible to apply a near-equilibrium approach to this system, in which Onsager coefficients and reciprocal relations are derived. This will be done in \ref{sec_rec}.

\section{Time-periodic thermodynamic uncertainty relations}\label{tTur}

Generically, { Thermodynamic Uncertainty Relations (TURs)} impose bounds between fluctuations
of a given flux and the associated dissipation rate. In its original formulation it states that for {an ergodic system described by a time independent Markov dynamics, the steady state currents (e.g. flux of particles or heat) obey a trade-off between signal and noise. The signal corresponds to the long-time limit of an
instantaneous current $\expval{j_t}$ -- that considers the jumps in time $t$ and their weights such as privileged directions in the state space -- averaged over all possible stochastic trajectories $\expval{j} =\lim_{t\to \infty} \expval{j_t}/t$. Conversely, the noise is given by its variance $\mathrm{Var}(j) = \lim_{t\to\infty} ( \expval{j^2_t} - \expval{j_t}^2)/t$  and the dissipation can be identified with the long-time limit of the instantaneous entropy production rate $\sigma = \lim_{t\to\infty} \Sigma_t/t$.

The existence of a universal bound relating above observables was conjectured by Barato et al. \cite{barato2015thermodynamic} and afterwards proven from the tools of large deviation theory \cite{touchette2009large} by Gingrich et al. \cite{gingrich2016dissipation} and Pietzonka et al. \cite{pietzonka2016universal}. It states that
\begin{equation}
    \frac{\sigma \mathrm{Var}(j) }{2k_B \left\langle j\right\rangle^2} \geq 1.\label{turor}
\end{equation}

Several works  \cite{holubec2018cycling, proesmans2017discrete, koyuk2018generalization, proesmans2019hysteretic, barato2019unifying, koyuk2019operationally, koyuk2020thermodynamic} have shown that this bound no longer holds in the presence of time-dependent driving and  hence Eq.~(\ref{turor}) generally 
will not be valid for the electron pump discussed previously. For this
reason, we will look at extensions  of Eq.~(\ref{turor}). In the next subsections we provide a concise overview of them. Although these bounds hold for any flux in the system, we shall cast them in terms of the particle fluxes into the reservoirs, cf.~Eqs.~(\ref{mc}) and (\ref{var}).
 }
 
\subsection{Effective entropy production relation}\label{efftur}

A first extension was proposed by Koyuk et al.~in Ref.~\cite{koyuk2018generalization} 
 and corresponds to a generalization of the original TUR \cite{barato2015universal}:
 \begin{equation}
	\frac{\textrm{Var}(J_i)\overline{\Pi}_\text{eff}}{2 \overline{J}_i^2} \geq 1,\label{tur2}
\end{equation}
where it takes into account the thermodynamic cost of the periodic driving of the system by a small modification in the standard entropy production rate, namely the effective entropy production $\overline{\Pi}_\text{eff}$.  In the case of the electron pump, it is given by

\begin{equation}
	\overline{\Pi}_\text{eff} { \equiv}	\frac{1}{\tau} \sum_{i=1}^N \int\limits_{(i-1)\tau'}^{i\tau'} \frac{(\omega_i - (\omega_i+\overline{\omega}_i)p^\text{eff})^2}{\omega_i - (\omega_i+\overline{\omega}_i) p^\text{ss}(t)} 
	 \ln \left[ \frac{(1- p^\text{ss}(t)) \omega_i}{ p^\text{ss}(t) \overline{\omega}_i}\right] \textrm{d}{t},
\end{equation}
where in the present case $p^\text{eff} { \equiv} \tau^{-1} \int_{0}^{\tau} p^\text{ss} (t) \textrm{d}{t} $ is the average occupancy of the quantum dot { over one steady-state cycle. Note that above expression reduces to the Eq. (\ref{sch}) 
when the steady-state is time-independent $p^\text{ss}(t)=p^\text{eff}$.}

\subsection{Generalized entropy production relation}
 
 A number of different generalizations of the TUR for time-periodic systems have been proposed by Barato et al.~\cite{barato2019unifying}. { Their validity depend on individual properties such as the time-dependence and symmetries presented in the increment of the current. Particularly suited for the present model is the  TUR given by (cf. GTUR4 in \cite{barato2019unifying})
 \begin{equation}
  \frac{\textrm{Var}(J_i)}{2 \overline{J_i}^2} C_a(p^\text{eff})\ge 1,\label{tur1}
 \end{equation}
that has the form of Eq.~(\ref{turor}) and for the present model
\begin{equation}
    C_a (p)  \equiv \frac{2}{\tau} 
\sum_{i=1}^N \int\limits_{(i-1)\tau'}^{i\tau'} 
    \frac{[(1-{ p})\omega_i - { p} \overline{\omega}_i ]^2}
    {(1-{ p^\text{ss}(t)})\omega_i - { p^\text{ss}(t)} \overline{\omega}_i} 
    \textrm{d}{t},\label{Cdef}
\end{equation}
where { $p$} is a  number between $0$ and $1$. Here, we set $p{ \to} p^\text{eff} \equiv \tau^{-1} \int_{0}^{\tau} p^\text{ss} (t) \textrm{d}{t} $ as introduced in Section \ref{efftur}. This relation is generally tighter  than Eq.~(\ref{tur2}) \cite{barato2019unifying}.}

\subsection{Hysteretic relation}
{
The third TUR to be considered was derived in \cite{proesmans2019hysteretic}, it relates the fluxes of the system in "forward" and "backward" (time-inverted driving) directions.
More specifically, it takes into account
the sum of the fluxes (and their variances) under forward  and time-reversed drivings \cite{proesmans2019hysteretic,potts2019thermodynamic}. Such a ``hysteretic'' TUR satisfies a bound similar to the one derived for time-symmetric driving \cite{proesmans2017discrete} and reads
\begin{equation}
  \frac{ \left(\textrm{Var}(J_i)+ \textrm{Var}(\tilde{J}_i)\right) \left( e^{\tau(\overline{\Pi} + \widetilde{\overline{\Pi}})/2k_B} -1\right)}{ \tau ( \overline{J_i} + \widetilde{\overline{J_i}} )^2}\ge 1,\label{tur3}
\end{equation}
where   $\widetilde{\overline{J_i}}$, $\widetilde{\overline{J^2_i}}$ and $\widetilde{\overline{\Pi}}$ stand for the results under time-inverted driving,
which can be appraised in terms of transition rates
\begin{equation}
    \tilde{\omega}(t) { \equiv} \omega(\tau-t)\qquad {\rm and} \qquad \tilde{\overline{\omega}}(t) { \equiv} \overline{\omega}(\tau-t).
\end{equation}
In our case, it corresponds to inverting the sequence of reservoirs to which the system is coupled.}

\subsection{Driving frequency relation}

The fourth and last TUR  we shall look at is a bound 
introduced by Koyuk et al.~\cite{koyuk2019operationally}, also referred as driving frequency relation, since  it takes into account the { response of the current to a change in the driving period}.
    More specifically the driving frequency TUR states that
\begin{equation}
 \frac{\overline{\Pi} \textrm{Var}(J_i)}{2(\overline{J_i} + \tau \frac{\partial \overline{J_i}}{\partial \tau} )^2}\ge 1.
\label{ksTUR}
\end{equation}
One advantage of this TUR is that it consists of quantities with clear macroscopic interpretations, that can be directly evaluated from the (forward) time dynamics.

\section{Comparing time-periodic thermodynamic uncertainty relations}\label{compar}
\begin{center}
\begin{figure}[h]
\includegraphics[width=.9\textwidth]{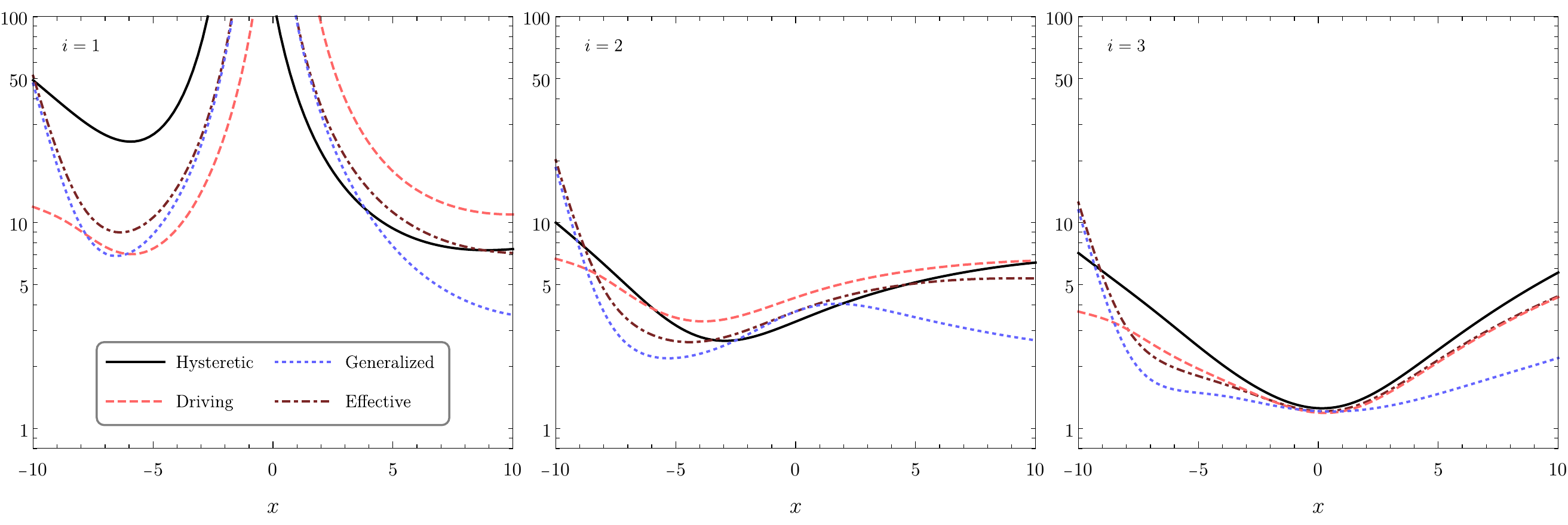}
\caption{Curves represent the LHS of TURs in Eqs.~(\ref{tur2}) (brown dot-dashed line), (\ref{tur1}) (blue dotted line),  (\ref{tur3}) (black full line) and (\ref{ksTUR}) (pink dashed line) for the electron flux $J_i$, associated with reservoir $i$ ($i=1,2,3$). These quantities need to be $\geq 1$ to satisfy each TUR. We consider the three stage electron pump versus parameter $x$ with affinities { $X_1=0$}, $X_2 = 0.2 x$ and $X_3 = -0.5x$. For $x = 0$ the system is at the equilibrium regime. Other parameters: $\omega_1=\overline{\omega}_1=0.3$, $\overline{\omega}_2=0.8$, $\overline{\omega}_3=0.2$ and $\tau=1$. Except for $i=1$, $\omega_i$, $\overline{\omega}_i$ and $X_i$ are related from Eq. (\ref{xidef}).}
\label{compar4}
\end{figure}
\end{center}

\begin{center}
\begin{figure}[h]
\includegraphics[width=.9\textwidth]{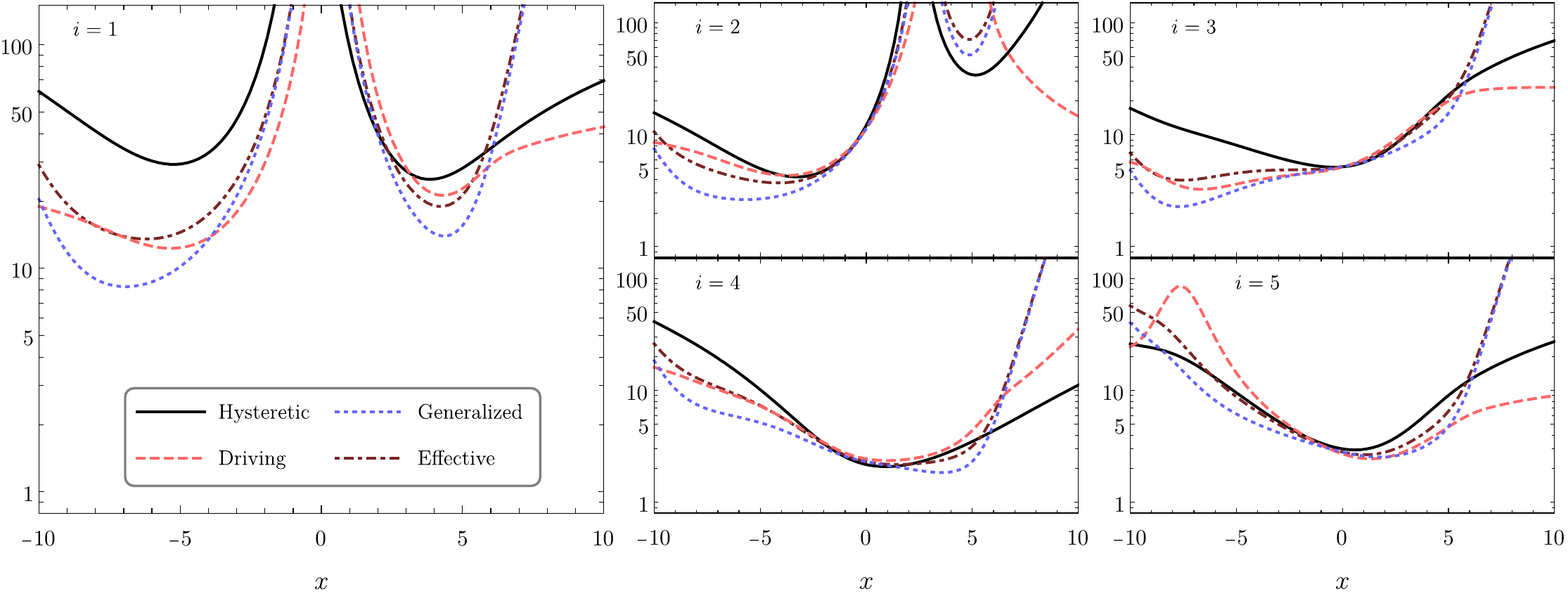}
\caption{Curves represent the LHS of TURs in Eqs.~(\ref{tur2}) (brown dot-dashed line), (\ref{tur1}) (blue dotted line),  (\ref{tur3}) (black full line) and (\ref{ksTUR}) (pink dashed line) for the electron flux $J_i$, associated with reservoir $i$ ($i=1,2,3,4,5$). These quantities need to be $\geq 1$ to satisfy each TUR. We consider the five stage electron pump versus parameter $x$ with affinities { $X_1=0$}, $X_2 = 0.2x$, $X_3 = -0.5x$, $X_4 = 0.7 x$ and $X_5 = -0.4x$. For $x = 0$ the system is at the equilibrium regime. Other parameters: $\omega_1=\overline{\omega}_1=0.3$, $\overline{\omega}_2=0.8$, $\overline{\omega}_3=0.2$, $\overline{\omega}_4=0.3$, $\overline{\omega}_5=0.5$ and $\tau=1$. Except for $i=1$, $\omega_i$, $\overline{\omega}_i$ and $X_i$ are related from Eq. (\ref{xidef}).}
\label{compar1}
\end{figure}
\end{center}

\begin{center}
\begin{figure}[h]
\includegraphics[width=.95\textwidth]{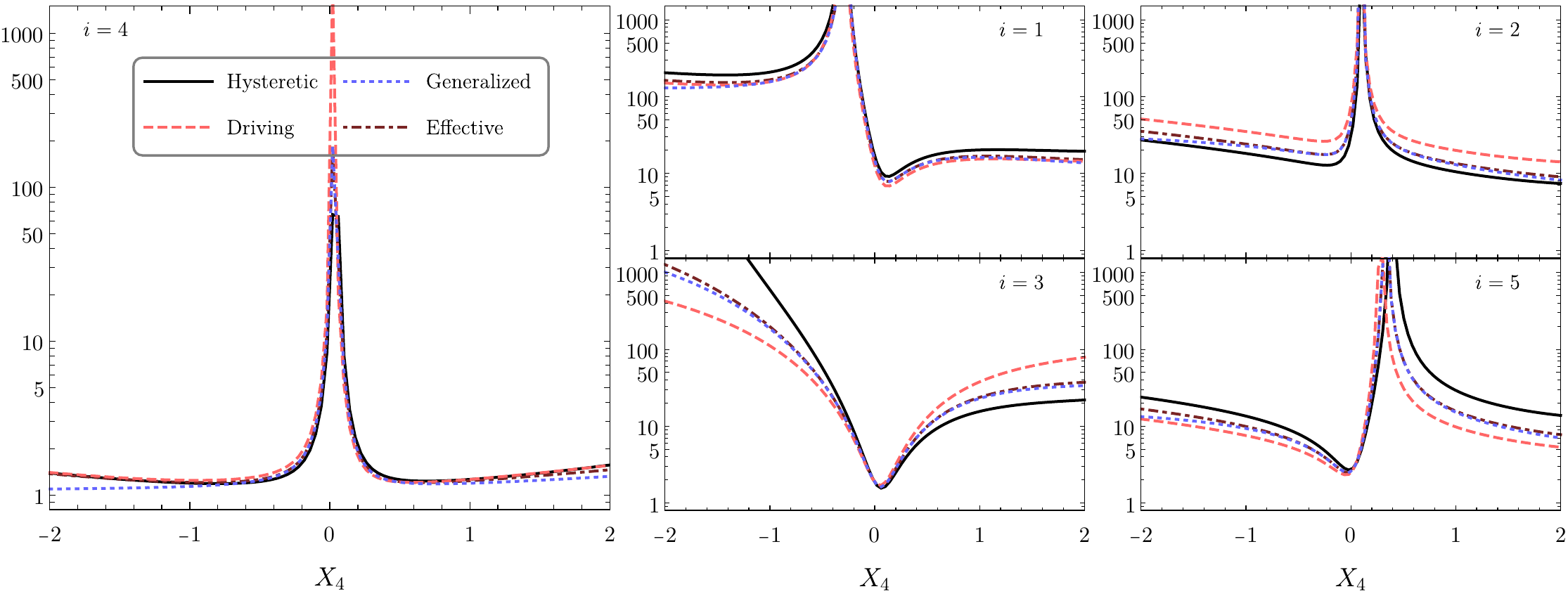}
\caption{Curves represent the LHS of TURs in Eqs.~(\ref{tur2}) (brown dot-dashed line), (\ref{tur1}) (blue dotted line),  (\ref{tur3}) (black full line) and (\ref{ksTUR}) (pink dashed line) for the electron flux $J_i$, associated with reservoir $i$ ($i=1,2,3,4,5$). These quantities need to be $\geq 1$ to satisfy each TUR. We consider the five stage electron pump versus  the affinity related to the fourth reservoir $X_4$. The others are $X_1=0$, $X_2 = 0.05$, $X_3 = -0.1$ and $X_5 = 0.1$. { The larger panel depicts the TUR associated with the fourth reservoir. Far from the origin the system is in the regime of one of the affinities being much larger than the others.} Also, we considered $\omega_1=\overline{\omega}_1=0.3$, $\overline{\omega}_2=0.8$, $\overline{\omega}_3=0.2$, $\overline{\omega}_4=0.3$, $\overline{\omega}_5=0.5$ and $\tau=1$. Except for $i=1$, $\omega_i$, $\overline{\omega}_i$ and $X_i$ are related from Eq. (\ref{xidef}). }
\label{compar3}
\end{figure}
\end{center}

\begin{center}
\begin{figure}[h]
\includegraphics[width=.95\textwidth]{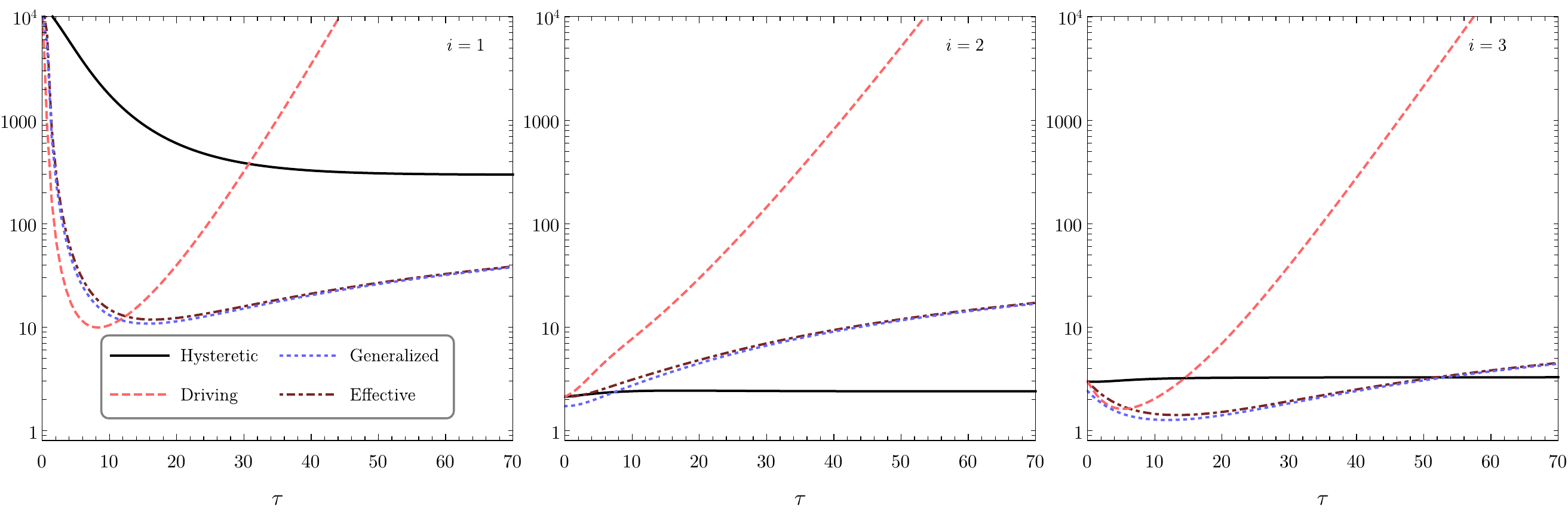}
\caption{Curves represent the LHS of TURs in Eqs.~(\ref{tur2}) (brown dot-dashed line), (\ref{tur1}) (blue dotted line),  (\ref{tur3}) (black full line) and (\ref{ksTUR}) (pink dashed line) for the electron flux $J_i$, associated with reservoir $i$ ($i=1,2,3$). These quantities need to be $\geq 1$ to satisfy each TUR. We consider the three stage electron pump versus period $\tau$ with affinities { $X_1=0$}, $X_2 = -2$ and $X_3 = 1.25$. Other parameters: $\omega_1=\overline{\omega}_1=0.7$, $\overline{\omega}_2=0.3$ and $\overline{\omega}_3=0.1$. Except for $i=1$, $\omega_i$, $\overline{\omega}_i$ and $X_i$ are related from Eq. (\ref{xidef}). }
\label{compar5}
\end{figure}
\end{center}

\begin{center}
\begin{figure}[h]
\includegraphics[width=.95\textwidth]{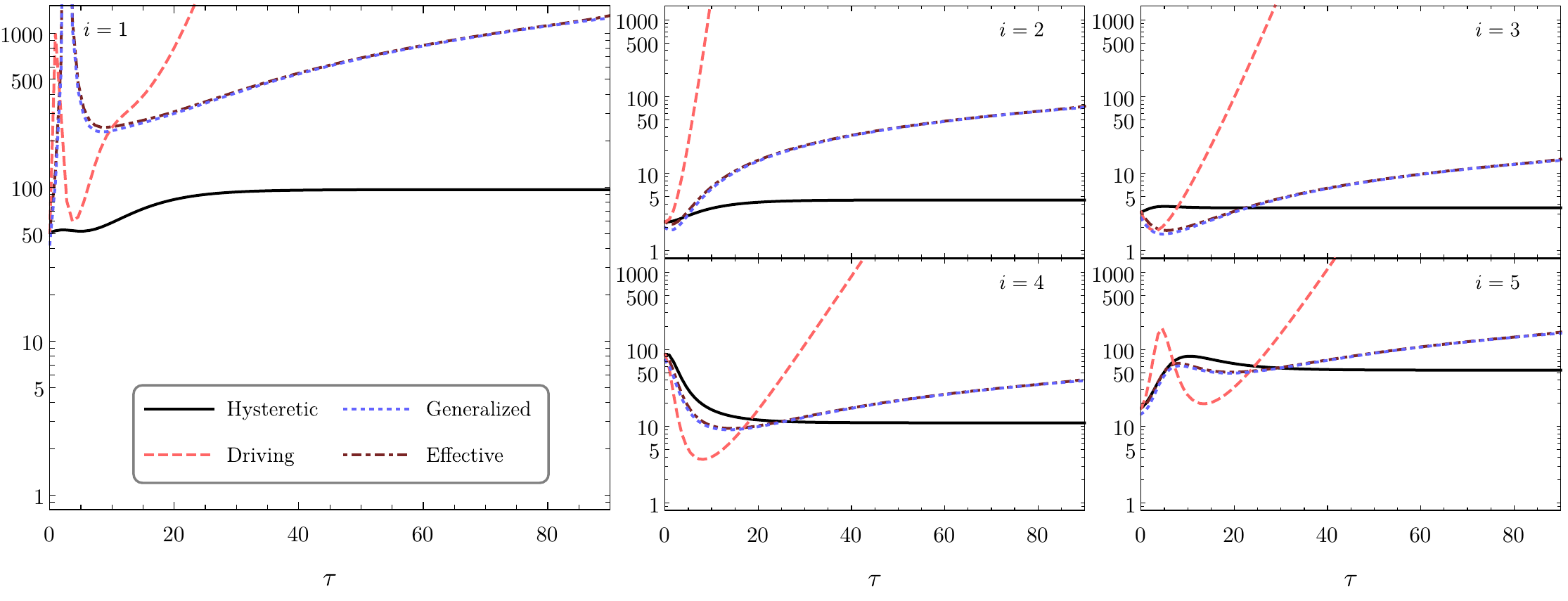}
\caption{Curves represent the LHS of TURs in Eqs.~(\ref{tur2}) (brown dot-dashed line), (\ref{tur1}) (blue dotted line),  (\ref{tur3}) (black full line) and (\ref{ksTUR}) (pink dashed line) for the electron flux $J_i$, associated with reservoir $i$ ($i=1,2,3$). These quantities need to be $\geq 1$ to satisfy each TUR. We consider the five stage electron pump versus period $\tau$ with affinities { $X_1=0$}, $X_2 = 1.8$, $X_3 = -1.5$, $X_4 = 0.7$, $X_5 = -0.4$, $\omega_1=\overline{\omega}_1=0.7$, $\overline{\omega}_2=0.3$, $\overline{\omega}_3=0.8$, $\overline{\omega}_4=0.2$ and $\overline{\omega}_5=0.5$. Except for $i=1$, $\omega_i$, $\overline{\omega}_i$ and $X_i$ are related from Eq. (\ref{xidef}).}
\label{compar42}
\end{figure}
\end{center}

{
We are now in position to apply the aforementioned TURs to the present system. The analysis will be divided in three parts: firstly, in Figs.~\ref{compar4} and \ref{compar1}, we shall compare them for fixed periods and distinct model parameters for the three and five stage cases.
We will scale all thermodynamic forces $X_i$'s with  a control parameter $x$. This gives us the possibility to study the near equilibrium regime when $x\approx 0$ and the far from equilibrium regime, $x\rightarrow\pm\infty$. Secondly,  Fig.~\ref{compar3} depicts, for the five stage case, the situation where one thermodynamic force is much smaller/larger than the others. More precisely, we keep all thermodynamic forces apart from $X_4$ fixed and analysis is undertaken for distinct values of $X_4$. Finally, in Figs.~\ref{compar5} and \ref{compar42}, we study the dependency of the TURs on the period of the driving. In all cases, we shall focus on the TURs applied to the particle fluxes $J_i$ into each reservoir $i$.}

In Figs.~\ref{compar4} and \ref{compar1}, one can see that the TURs are indeed valid, as expected. Furthermore, they seem to converge to a similar value near equilibrium, $x\approx 0$, although divergences might occur when the denominators in the TUR go to zero (for example when the average flux $\overline{J}_i$ vanishes). 
 Furthermore, for most fluxes, the TURs seem more tight near equilibrium. One can also easily check that, although the TURs seem to have a qualitatively similar behaviour over a broad range of parameters there is no TUR which is uniformly better than the others. The behavior of an electron pump with three reservoirs, Fig.~\ref{compar4}, is qualitatively similar to the behaviour of the electron pump with five reservoirs,  Fig.~\ref{compar1}. Finally, one can see that the correct choice of flux is important if one wants to infer the entropy production rate from the TUR. For example, in Fig.~\ref{compar4}, one can see that the bound for $J_3$ is reasonably tight in general, but the bound for $J_1$ is off by a factor $100$ over a broad range of parameters.

This can also be seen in Fig.~\ref{compar3}, where one only varies $X_4$. If $X_4$ is small, the TURs associated with $J_4$ are generally less tight, while the TURs associated with the other reservoirs are tighter. Meanwhile, the TURs associated with $J_4$ seem to become tighter for larger $X_4$. The reason for this is that for large $X_4$, the entropy production is dominated by the contact with the $4$-th reservoir. Therefore, one can get a good estimate of the total entropy production, or a tight thermodynamic uncertainty relation, by just looking at the $4$th flux. On the other hand, all other fluxes become small compared to the entropy production, meaning that the TURs associated to the other fluxes become loose.

Finally in Figs.~\ref{compar5} and \ref{compar42}, we study the dependency of the TURs on the period of the driving. From expressions of $\overline{J_i}$, $\textrm{Var}(J_i)$, $\overline{\Pi}$, $\overline{\Pi}_\text{eff}$ and $C_a(p_\text{eff})$, it is possible to find
the asymptotic behaviors of TURs
for sufficiently slow and fast periods. In fact they acquire the asymptotic behaviors for large $\tau$
\begin{equation}
\overline{J}_i\sim \frac{1}{\tau}, \quad \textrm{Var}({J}_i)\sim \frac{1}{\tau}, \quad \overline{\Pi}\sim \frac{1}{\tau},\quad \overline{\Pi}_{\textrm{eff}}\sim 1,\quad  C_a(p^{\textrm{eff}})\sim 1, \quad \ln\left|\overline{J_i} + \tau \frac{\partial \overline{J_i}}{\partial \tau}\right|\sim -\tau.
\end{equation}
This implies that the TURs (\ref{tur2}) and (\ref{tur1}) diverge linearly and Eq.~(\ref{ksTUR}) exponentially as $\tau$
is increased, whereas the hysteretic TUR, Eq.~(\ref{tur3}), remains finite. Conversely, for $\tau \rightarrow 0$ all of above quantities remain finite, hence all TURs are limited in the slow driving regime, even though they can take very large values.

\section{Conclusions}\label{conc}
We studied the nonequilibrium properties of a single-particle stochastic pump sequentially placed in contact with an arbitrary number of reservoirs. Exact expressions for fluxes and thermodynamic quantities  are found arbitrarily far from equilibrium. We tested and verified several TURs for time-periodic systems. Near equilibrium, all TURs yield similar results, but far from equilibrium their tightness often differ by several orders of magnitude. Furthermore, there is no TUR that seems to be uniformly better than the others. One can however verify that in the slow-driving limit all TURs, except for the hysteretic TUR Eq.~(\ref{tur3}), become arbitrarily loose. It would be interesting to check whether this conclusion is true for more general periodically driven systems. Inferring quantities such as the average flux or the entropy production rate from these TURs seems inaccurate, since TURs for individual fluxes often differ from their lower bounds by orders of magnitude. It would be interesting to see whether one can overcome this issue by looking at combined fluxes, which might lead to hyperaccurate currents for which the TUR becomes tight \cite{busiello2019hyperaccurate}. 

\section{Acknowledgements}
C. E. F. and P. E. H. acknowledge the financial support from FAPESP
under grants No 2018/02405-1 and 2017/24567-0, respectively.

\appendix

\section{Linear regime and Onsager coefficients}\label{sec_rec}
\begin{figure}[h]
\includegraphics[width=\textwidth]{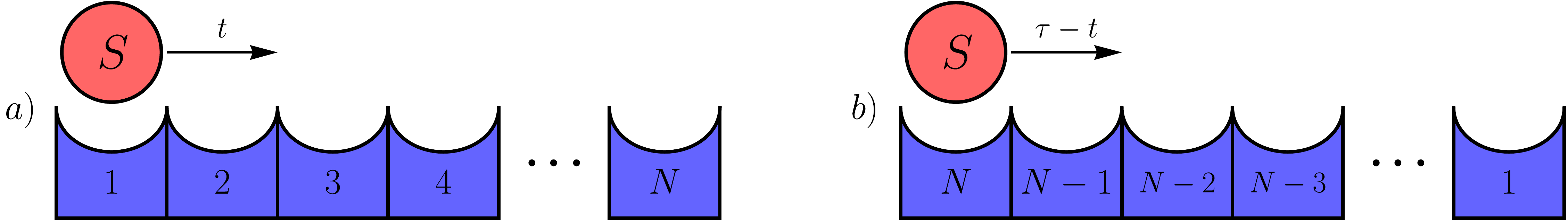}
\caption{Illustration of the sequence of reservoirs that interacts with the system at each interval $\tau/N$ in a cycle. a) is the forward protocol while b) is the time-reversed one.}
\label{trajs}
\end{figure}

In this appendix, we derive the reciprocal relations for the system exposed to an arbitrary number of sequential reservoirs.
{
In the absence of odd parity variables,  Onsager coefficients generally satisfy
the so called Onsager reciprocal relations, $L_{ij}=L_{ji}$.
This is no longer the case for systems with time-dependent drivings, since
 the time-reversal symmetry is broken in such a case. For continuous time periodic drivings, Onsager
symmetry is replaced by the weaker  Onsager-Casimir reciprocal relation,
which relates the Onsager coefficients under time-forward
driving to the cross-coefficient of time-reversed driving,
\begin{equation}
  {\it L}_{ij}={\it \tilde{L}}_{ji},
  \label{casimir}
\end{equation}
where $\tilde{L}$ attempts to the reversed driving.
 Since we are dealing with a distinct type of periodically driven system, 
 Eq. (\ref{casimir}) has to  be updated in order
 to incorporate the sequential exposure and hence a different structure for reciprocal relations is expected.}
From the linear expansion
of the mean current ${\bar J}_i$ [Eq. (\ref{mc})], we have that $L_{i,j}$ reads  
\begin{equation}
	L_{i,j} = \frac{\partial \overline{J}_i}{\partial X_j}\biggr\vert_{\text{eq}}= \nu_j \frac{\partial \overline{J}_i}{\partial \nu_j} \biggr\vert_{\text{eq}},
\end{equation}
where  we introduced the quantity $\nu_i = \omega_i/\overline{\omega}_i$ and the subscript ``eq" refers to the equilibrium case in which $X_i =0$ (or $\nu_i=\nu$, $\forall i$).
For $j\leq i$ and $m=\{1,2,\dots ,N\}$ they are then given by

\begin{align}
	L_{i,j}= &\frac{ (1-\xi_{j,j}) (\xi_{1,i-1} \xi_{j+1,N} +\xi_{j+1,i-1}(1-\xi_{1,N})) + \delta_{i,j} (1- \xi_{1,N})(\xi_{j,j}-2)  }{1-\xi_{1,N}} \times \nonumber \\
	&\times \frac{(\xi_{i,i}-1) \nu_j}{ \tau (\nu_j+1)^2}\biggr\vert_{\nu_m\to\nu}
	\label{jlei}
\end{align}
and for $j>i$ and $m=\{1,2,\ldots ,N\}$ they are
\begin{equation}\label{ilj}
	L_{i,j}=\frac{\xi_{1,i-1}(1-\xi_{j,j})\xi_{j+1,N}}{1-\xi_{1,N}} \frac{(\xi_{i,i}-1) \nu_j}{ \tau (\nu_j+1)^2}\biggr\vert_{\nu_m\to\nu}.
\end{equation}
Above expressions are held valid for an arbitrary number
of reservoirs $N$.
Taking for instance the forward protocol and its reverse order as sketched in Fig. \ref{trajs}, the  time reversal consists  of inverting the order of reservoir exposures.
Since the process is cyclic,  it can be accomplished by taking the backward sequence $\{N,N-1,\ldots,1\}$, in which  the $N$-th reservoir  plays the role of reference in the reversed dynamics, so the equilibrium probability of occupation is now given by $p_N^\text{eq}$.

Thereby, the  reciprocal relation
for an arbitrary number of sequential reservoirs reads
\begin{equation}\label{recip}
	L_{i,j}= \tilde{L}_{\tilde{j},\tilde{i}},
\end{equation}
with $\tilde{i}=N+1-i$, $\tilde{j}=N+1-j$, the $i$-th and $j$-th reservoir under time-inverted driving, for $i,j=1,...,N$. The Onsager matrix is symmetric in relation to the Onsager matrix for the time-reversed dynamics.

The time reversal symmetry then corresponds
to perform the transformation {$i\to \tilde{j}$} and {$j\to \tilde{i}$} and hence the function $\xi_{i,j}$ becomes
{
\begin{align}
	\xi_{i,j}&\to \xi_{\tilde{j},\tilde{i}}.
\end{align}
}
For $j<i$ the right side in Eq. (\ref{jlei}) reads
\footnotesize{
\begin{align}
	{\tilde{L}_{\tilde{j},\tilde{i}}}&= \biggr[ \frac{ (1-\xi_{N-i+1,N-i+1})( \xi_{N-i+2,N} \xi_{1,N-j} + \xi_{N-i+2,N-j} (1- \xi_{1,N}))}{1-\xi_{1,N}} \times \nonumber \\
	&\hspace{.7cm}\times\frac{(\xi_{N-j+1,N-j+1}-1)\nu_{N-i+1}}{ \tau (\nu_{N-i+1}+1)^2} \biggr]_\text{eq}^\sim \\
	&= \frac{(1-\xi_{i,i})( \xi_{j+1,N} \xi_{1,i-1}+ \xi_{j+1,i-1}(1-\xi_{1,N}))}{1-\xi_{1,N}} \frac{(\xi_{j,j}-1)\nu}{\tau (\nu+1)^2}\biggr\vert_\text{eq} = L_{i,j}.
\end{align}}

\normalsize
Analogous result can be obtained for  $i\leq j$. With these results for { $\tilde{L}_{\tilde{j},\tilde{i}}$} in the cases $j<i$ and $i\leq j$, \eqref{recip} is analytically verified.
This then completes our proof about the appropriate reciprocal relations for the present case. The above result is valid for an arbitrary number of reservoirs.

\section*{Bibliography}
\providecommand{\newblock}{}

\end{document}